\documentclass{aa}

\usepackage{graphicx}
\usepackage[varg]{txfonts}
\usepackage{comment}
\usepackage{soul}
\usepackage{tabularx}
\usepackage{soul}
\usepackage{natbib}
\usepackage{hyperref}
\usepackage{cleveref}
\usepackage{siunitx}
\usepackage{tabularx}
\usepackage{xcolor}
\usepackage{float}
\usepackage{bm}
\usepackage{gensymb}
\usepackage[normalem]{ulem}
\usepackage[caption=false]{subfig}
\usepackage{txfonts}
\usepackage{diagbox}
\usepackage{soul}
\usepackage{multirow}
\usepackage{comment}
\usepackage{ulem}

\begin{document}

\title{Decrease in Milky Way rotation curve revisited}

   \author{J.~Kla\v{c}ka
          \inst{1}
          \and
           M. \v{S}turc \inst{1}
                    }

   \institute{Faculty of Mathematics, Physics, and Informatics, Comenius University, 
	Mlynsk\'{a} dolina, 842 48 Bratislava, Slovak Republic\\
              \email{jozef.klacka@fmph.uniba.sk,  michal.sturc@fmph.uniba.sk}
             }

   \date{Received 2025}

 \abstract
   {Latest papers on the rotation curve of the Milky Way galaxy, i. e. \citet{Ou, Jiao, 2023ApJ...945....3S} suggest a Keplerian decrease in the rotation curve. This behavior is not consistent with other spiral galaxies \citep{SPARCmaster,Mistele2024}.}
   {Show that the prior use of the axisymmetric Jeans equation is not consistent with the final model produced in the papers.}
   {Comparison of the results on gravitational potential in \citet{Ou, Jiao, 2023ApJ...945....3S} with the prior assumptions about the axisymemtric properties of the Milky Way galaxy.}
   {The gravitational potentials published by \citet{Ou, Jiao, 2023ApJ...945....3S} lead to almost spherically symmetric properties of the Milky Way galaxy at Galactocentric radii above $20~\mbox{kpc}$, which is not consistent with the use of axisymmetric Jeans equations.}
{}


   \maketitle

\section{Introduction}\label{introduction-N}

Recently published papers, i.e. \citet{2023ApJ...945....3S,Jiao,Ou} suggest that the Milky Way (MW) circular velocity curve significantly differs from the rotation curves of other spiral galaxies. The new results suggest that the MW rotation curve decreases with the Galactocentric distance $R$ in a Keplerian manner. As a consequence, the amount of dark matter (DM) in the MW galaxy is significantly lower than was assumed. Moreover, the result for the MW galaxy differs from the result found for other spiral galaxies, see e.g. \citet{SPARCmaster} or \citet{Mistele2024}.

Why does mass distribution of our galaxy differ from the distribution of other spiral galaxies?

The papers  \citet{2023ApJ...945....3S,Jiao,Ou} 
obtain the MW rotation curve from the Jeans equations for systems with axisymmetric cylindrical mass distribution. However, the distribution of matter at large Galactocentric distances is mostly spherical. Furthermore, the MW galaxy is warped, which invalidates certain assumptions about the symmetry of the Galactic disk.

\section{Published decrease of the rotation curve of the MW}

The decrease in the circular velocity curve of the MW galaxy, $v_{c}(R)$, is very slight for distances approximately between $8~\mbox{kpc}$ and $15~\mbox{kpc}$. However, for larger values of $R$, the decrease in $v_{c}(R)$ becomes more significant, see i.e. \citet{2023ApJ...945....3S,Jiao,Ou}. For example, the newest paper by \citet{Ou} presents $v_{c}(R = 27.3~\mbox{kpc}) = (173.0 \pm 17.1) ~\mbox{km}\mbox{s}^{-1}$, which is considerably lower than the circular velocity at the Sun's distance, where $v_{c} \approx 230~\mbox{km}\mbox{s}^{-1}$.

\section{Discrepancy between published accelerations and theoretical approach}

The inner part of the MW galaxy disk (approximately $5~\mbox{kpc} < R < 10~\mbox{kpc}$) can be accurately described by axisymmetric mass distribution. Thus, Jeans equations for axisymmetric systems can be used to interpret the observational data, see \citet[Eqs. 4.222, 4.226]{binney} and \citet[Eqs. 7, 8]{Ou}. However, for larger values of $R$, the spherically distributed DM becomes gravitationally dominant over the axisymmetric disk. Thus, the Jeans equations for spherical systems \citet[Eq. 4.214]{binney} should be used.

\subsection{Circular velocities} 
We can write $v_{BM}^{2} + v_{DM}^{2} = v_{c}^{2}$ for the circular velocities generated by the baryonic matter (BM) and dark matter (DM), $v_{c}$ is the real circular velocity corresponding to observations. Relative contribution of BM to the circular velocity curve is given by the ratio $v_{BM}^{2}/v_{c}^{2} \equiv \left [ v_{c}(R) \right ]^{2}_{BM}/ \left [ v_{c}(R) \right ]^{2}$ for any Galactocentric distance $R$.

\subsubsection{Circular velocity values published in \citet{Ou}}\label{c-v-Ou}
For the values used in the best-fitting model in \citet{Ou} $\left [ v_{c}(R=27.3~\mbox{kpc}) \right ]^{2}_{BM}/ \left [ v_{c}(R=27.3~\mbox{kpc}) \right ]^{2} = 0.12 \pm 0.02$. The small value of $0.12$ confirms that the spherically symmetric DM distribution is dominant over the axisymmetric BM distribution for $R = 27.3~\mbox{kpc}$. 

The calculated value $v_{c}(R=27.3~\mbox{kpc}) = (173.0 \pm 17.1)~\mbox{km} ~\mbox{s}^{-1}$ follows from the Jeans equations valid for the axisymmetric systems. Thus, the result $v_{c}(R=27.3~\mbox{kpc}) = (173.0 \pm 17.1)~\mbox{km} ~\mbox{s}^{-1}$ is not consistent with the prior assumption of the axisymmetric distribution. 

\subsubsection{Circular velocity values corresponding to a flat rotation curve}\label{c-v-flat}
For a hypothetical flat rotation curve with $v_{c} = 230~\mbox{km}\mbox{s}^{-1}$ (approximate value taken from \citet{Ou} for the region of the Sun), also $v_{c}(R=27.3~\mbox{kpc}) \doteq 230 ~\mbox{km} \mbox{s}^{-1}$, we obtain $\left [ v_{c}(R=27.3~\mbox{kpc}) \right ]^{2}_{BM}/ \left [ v_{c}(R=27.3~\mbox{kpc}) \right ]^{2} = 0.07$. The value $0.07$ confirms that the spherically symmetric DM distribution would still be dominant over the axisymmetric BM distribution for $R = 27.3~\mbox{kpc}$. 

\subsubsection{Circular velocity values and the Jeans equations for axisymmetric systems}  
The values found for $R = 27.3~\mbox{kpc}$ in Secs. \ref{c-v-Ou} and \ref{c-v-flat}, $0.12$ and $0.07$, confirm that the action of the spherically distributed DM is significant and the use of the Jeans equations for axisymmetric systems is not justified. 

\subsection{Comparison of accelerations}
The best-fitting values presented in \citet{Ou} (Tables 2 and 3) lead to a model with double exponential discs and an Einasto DM halo. In Table \ref{Table11111}, we summarize the accelerations in the radial direction caused by spherically symmetric MW components (central bulge and DM halo) and axisymmetric components (stellar and gaseous discs). Because the percentage of acceleration caused by the axisymmetric components drops below $20\%$, we can conclude that the MW galaxy at this distance behaves as a spherically symmetric system. This is, however, inconsistent with the use of the axisymmetric Jeans equation by the authors (Eqs. 7 and 8).

\begin{table}[h]
\caption{\centering{{Radial accelerations implied by the best-fitting model in \cite{Ou}}}}
\centering
\def\arraystretch{1.3}
\begin{tabular}{ |c|c|c|c| } 
 \hline
 $R~[\mbox{kpc}]$ & $g_{sph}$ & $g_{cyl}$ & $\%_{sph}$ \\
 \hline
 19.71 & 1711 & 486 & 77.88 \\
 20.22 & 1652 & 461 & 78.19 \\
 20.72 & 1597 & 438 & 78.47 \\
 21.22 & 1544 & 417 & 78.73 \\
 21.72 & 1494 & 398 & 78.96 \\
 22.27 & 1440 & 378 & 79.19 \\
 22.71 & 1399 & 364 & 79.37 \\
 23.40 & 1337 & 342 & 79.61 \\
 25.02 & 1204 & 300 & 80.01 \\
 27.31 & 1044 & 253 & 80.52 \\
 \hline
\end{tabular}
\tablefoot{Table shows the radial gravitational acceleration caused by the spherically symmetric components ($g_{sph}$), axisymmetric components ($g_{cyl}$) and the percentage of total radial acceleration caused by the spherically symmetric components ($\%_{sph}$). All accelerations are in units of $\mbox{km}^{2}~\mbox{s}^{-2} ~\mbox{kpc}^{-1}$. The model values are consistent with observational values presented by \cite{Ou}, where the model values are calculated as $v_c(R) = \sqrt{R \left ( g_{sph} + g_{cyl} \right )}$. The value of $v_{c}$ for the most distant stars is $\left [ v_{c}(R=27.3~ \mbox{kpc}) \right ]_{model} = 188.2~\mbox{km}~\mbox{s}^{-1}$, which is consistent with the observational result $(173.0 \pm 17.1)~\mbox{km}~\mbox{s}^{-1}$.}
\label{Table11111}
\end{table}

\subsection{Jeans equations and spherically symmetric systems}  
If we make an approximation $\langle v_{\theta}^{2} \rangle = \langle v_{\phi}^{2} \rangle$ in \citet[Eq. 4.214]{binney}, then comparison with \citet[Eq. 4.226]{binney} and \citet[Eqs. 7, 8]{Ou} leads to 

\begin{equation}\label{approx}
    [v_{c}(R)]_{spherical} \doteq \sqrt{2} ~[v_{c}(R)]_{axisymmetric}
\end{equation}
for a distance $R$ where the spherically distributed dark matter is dominant.
We can remind that Eq. (8) in \citet{Ou} does not consider $v_{z}$, the $z-$component of the velocity. Real values of $|v_{z}|$ may be greater than $100~\mbox{km}~\mbox{s}^{-1}$, see the data cut-off in \citet{Ou}.

\section{Application}
If the Jeans equation for spherical systems is used, then the approximation presented in Eq. (\ref{approx}) can be used and it yields $v_{c}(R=27.3~\mbox{kpc}) \doteq \sqrt{2} (173.0 \pm 17.1)~\mbox{km}\mbox{s}^{-1} \doteq (244.7 \pm 24.2)~\mbox{km}~\mbox{s}^{-1}$. This value is consistent with a flat rotation curve of the MW galaxy.

Approximately for $5~\mbox{kpc} < R < 12~\mbox{kpc}$, ne can find rotation curve of the MW galaxy on the basis of the results presented by e.g. \citet{Ou} or \citet{2019ApJ...871..120E}.

\section{Milky Way warp}

Another reason why the cylindrical Jeans equations can't be used beyond $R \approx 20~\mbox{kpc}$ is that the Milky Way disk becomes warped with an amplitude of about $z_{warp} \approx (0.3-0.4)~\mbox{kpc}$ for stars younger than $5~\mbox{Gyr}$ \citep{Warp} and $z_{warp} \approx (1.3-1.5)~\mbox{kpc}$ for stars younger than $10~\mbox{Gyr}$ \citep{warp_new}. This means that stars with the $z$ coordinate equal to $0~\mbox{kpc}$ still experience non-zero gravitational acceleration in the $z$-direction. \cite{Simbad}

\section{Conclusion}
Recent papers lead to decreasing circular velocity curve for the Milky Way galaxy, see \citet{2023ApJ...945....3S,Jiao,Ou}. What is the reason for this decrease?

Our paper suggests that no physical process acting on the MW's matter generates the decrease. The decrease of the rotation curve is caused by incorrect data analysis.

The paper shows that it is not justified to use Jeans equations derived for the axisymmetric systems when working at large galactocentric distances in the MW galaxy. If we use the Jeans equations for the spherically symmetric systems, then we obtain the flat rotation curve for the MW galaxy. 

The result about the flat circular velocity curve for the MW galaxy is consistent with the rotation curves obtained for other spiral galaxies, see, e.g., \citet{SPARCmaster} and \citet{Mistele2024}.

\section*{Acknowledgement} This work was supported by the VEGA - the Slovak Grant Agency for Science, grant No. 1/0481/25.

\bibliography{Decrease.bbl}{}

\begin{thebibliography}{}
\expandafter\ifx\csname natexlab\endcsname\relax\def\natexlab#1{#1}\fi
\providecommand{\url}[1]{\href{#1}{#1}}
\providecommand{\dodoi}[1]{doi:~\href{http://doi.org/#1}{\nolinkurl{#1}}}
\providecommand{\doeprint}[1]{\href{http://ascl.net/#1}{\nolinkurl{http://ascl.net/#1}}}
\providecommand{\doarXiv}[1]{\href{https://arxiv.org/abs/#1}{\nolinkurl{https://arxiv.org/abs/#1}}}

\bibitem[{{Binney} \& {Tremaine}(2008)}]{binney}
{Binney}, J., \& {Tremaine}, S. 2008, {Galactic Dynamics: Second Edition} (Princeton University Press, Princeton)

\bibitem[{{Chrob{\'a}kov{\'a}} {et~al.}(2022){Chrob{\'a}kov{\'a}}, {Nagy}, \& {L{\'o}pez-Corredoira}}]{Warp}
{Chrob{\'a}kov{\'a}}, {\v{Z}}., {Nagy}, R., \& {L{\'o}pez-Corredoira}, M. 2022, \aap, 664, A58, \dodoi{10.1051/0004-6361/202243296}

\bibitem[{{Eilers} {et~al.}(2019){Eilers}, {Hogg}, {Rix}, \& {Ness}}]{2019ApJ...871..120E}
{Eilers}, A.-C., {Hogg}, D.~W., {Rix}, H.-W., \& {Ness}, M.~K. 2019, The Astrophysical Journal, 871, 120, \dodoi{10.3847/1538-4357/aaf648}

\bibitem[{{Jiao} {et~al.}(2023){Jiao}, {Hammer}, {Wang, Haifeng}, {Wang, Jianling}, {Amram, Philippe}, {Chemin, Laurent}, \& {Yang, Yanbin}}]{Jiao}
{Jiao}, Y., {Hammer}, F., {Wang, Haifeng}, {et~al.} 2023, \aa, 678, A208, \dodoi{10.1051/0004-6361/202347513}

\bibitem[{{Lelli} {et~al.}(2016){Lelli}, {McGaugh}, \& {Schombert}}]{SPARCmaster}
{Lelli}, F., {McGaugh}, S.~S., \& {Schombert}, J.~M. 2016, \aj, 152, 157, \dodoi{10.3847/0004-6256/152/6/157}

\bibitem[{{Mistele} {et~al.}(2024){Mistele}, {McGaugh}, {Lelli}, {Schombert}, \& {Li}}]{Mistele2024}
{Mistele}, T., {McGaugh}, S., {Lelli}, F., {Schombert}, J., \& {Li}, P. 2024, \apjl, 969, L3, \dodoi{10.3847/2041-8213/ad54b0}

\bibitem[{{Ou} {et~al.}(2024){Ou}, {Eilers}, {Necib}, \& {Frebel}}]{Ou}
{Ou}, X., {Eilers}, A.-C., {Necib}, L., \& {Frebel}, A. 2024, \mnras, 528, 693, \dodoi{10.1093/mnras/stae034}

\bibitem[{{Sylos Labini} {et~al.}(2023){Sylos Labini}, {Chrob{\'a}kov{\'a}}, {Capuzzo-Dolcetta}, \& {L{\'o}pez-Corredoira}}]{2023ApJ...945....3S}
{Sylos Labini}, F., {Chrob{\'a}kov{\'a}}, {\v{Z}}., {Capuzzo-Dolcetta}, R., \& {L{\'o}pez-Corredoira}, M. 2023, \apj, 945, 3, \dodoi{10.3847/1538-4357/acb92c}

\bibitem[{{Uppal} {et~al.}(2024){Uppal}, {Ganesh}, \& {Schultheis}}]{warp_new}
{Uppal}, N., {Ganesh}, S., \& {Schultheis}, M. 2024, \mnras, 527, 4863, \dodoi{10.1093/mnras/stad3525}

\bibitem[{{Wenger} {et~al.}(2000){Wenger}, {Ochsenbein}, {Egret}, {Dubois}, {Bonnarel}, {Borde}, {Genova}, {Jasniewicz}, {Lalo{\"e}}, {Lesteven}, \& {Monier}}]{Simbad}
{Wenger}, M., {Ochsenbein}, F., {Egret}, D., {et~al.} 2000, \aaps, 143, 9, \dodoi{10.1051/aas:2000332}

\end{thebibliography}
\bibliographystyle{aasjournal}

\end{document}